\documentclass[journal=jacsat,manuscript=article]{achemso}
\setkeys{acs}{articletitle = true}
\usepackage[version=3]{mhchem} % Formula subscripts using \ce{}
\usepackage[T1]{fontenc}       % Use modern font encodings
\usepackage{amssymb}
\author{Tokarchuk M.}
\email{mtok@icmp.lviv.ua}
\affiliation{Institute for Condensed Matter Physics of the National Academy of Sciences of Ukraine, \\ 1 Svientsitskii str., 79011, Lviv, Ukraine}
\altaffiliation{Lviv Polytechnic National University,  12 S. Bandera str., 79013, Lviv, Ukraine}

\title[To the kinetic theory of dense gases and liquids. Calculation of quasi-equilibrium particle \ldots]
  {To the kinetic theory of dense gases and liquids. Calculation of quasi-equilibrium particle distribution functions by the method of collective variables}

\abbreviations{IR,NMR,UV}

\begin{document}

%%%%%%%%%%%%%%%%%%%%%%%%%%%%%%%%%%%%%%%%%%%%%%%%%%%%%%%%%%%%%%%%%%%%%%
%%% The "tocentry" environment can be used to create an entry for the
%%% graphical table of contents. It is given here as some journals
%%% require that it is printed as part of the abstract page. It will
%%% be automatically moved as appropriate.
%%%%%%%%%%%%%%%%%%%%%%%%%%%%%%%%%%%%%%%%%%%%%%%%%%%%%%%%%%%%%%%%%%%%%%
%\begin{tocentry}
%
%Some journals require a graphical entry for the Table of Contents.
%This should be laid out ``print ready'' so that the sizing of the
%Inside the \texttt{tocentry} environment, the font used is Helvetica
%8\,pt, as required by \emph{Journal of the American Chemical
%The surrounding frame is 9\,cm by 3.5\,cm, which is the maximum
%permitted for  \emph{Journal of the American Chemical Society}
%graphical table of content entries. The box will not resize if the
%content is too big: instead it will overflow the edge of the box.
%
%This box and the associated title will always be printed on a
%separate page at the end of the document.
%
%\end{tocentry}

\begin{abstract}
  Based on a chain of BBGKI equations with a modified boundary condition
     that takes into account multiparticle correlations,
     kinetic equations in the approximate ``pairs'' collisions and in the polarization approximation,
     taking into account the interaction through the third particle,
     obtained.
     The specifics of the model representation of the pair potential of particle interaction through short-range and long-range parts were taken into account.
     In the case of the short-range potential in the form of the potential of solid spheres,
     the contribution of Enskog's revised theory to the complete integration of the collision of the kinetic equation is obtained.
     The collision integrals include paired quasi-equilibrium distribution functions that depend on the nonequilibrium mean values of the particle number density and the inverse temperature.
     The method of collective variables Yukhnovskii is applied for the calculation of pair quasi-equilibrium distribution function with an allocation of short-range and long-range parts in the potential of the interaction of particles.
     In this case,
     the system with short-range interaction is considered as a frame of reference.
\end{abstract}

\section{Introduction}
 Studies of the role of repulsive and attractive forces between atoms, molecules of liquids and their influence on thermodynamic,
 structural,
 and dynamic properties have a history,
 starting with Van der Waals~\cite{Waals,Zwanzig,Rowlinson,Widom,Fisher,Widom11,Longuet,Barker,Weeks,Kushickt,Chandler,Barker1,Kushickt1,Sung,Barrat,
Chandler1,Hansen,KirchnerB,Berthier,Melnyk,Pedersen,Tox1,Tox2,Berthier1,Ingebrigtsen,Hordiichuk}.
 And this is probably not all the work in which the problem is primarily the correct definition of the nature of forces,
 the construction of model capacity within the first coordination sphere of the liquid state.
 Everything that happens with the nature of the interaction of particles in the first coordination sphere of liquids is essential because,
 for highly correlated liquids,
 the contribution of long-range attraction will not be significant.
 For complex fluids:
 molecular,
 polar,
 and magnetic,
 when the orientational degrees of freedom are also necessary,
 the structure of the first coordination sphere will obviously determine the nature of the forces of interaction,
 the emergence of certain cluster formations,
 collective hydrodynamic sound,
 thermal modes,
 and nonhydrodynamic modes~\cite{Mryglod1,Mryglod2,Omelyan,Markiv}.
 On a spatial and temporal scale,
 nonequilibrium processes,
 which are the main within the first coordination sphere of simple fluids,
 belong to the kinetic stage of the temporal evolution of transfer processes.

 In the nonequilibrium theory of dense gases and liquids,
 the approach of kinetic equations makes it possible to obtain the transfer equation for the hydrodynamic densities of the number of particles,
 their momentum,
 and energy,
 and the expressions for the transfer coefficients as a function of the particle parameters and external parameters~\cite{Bogolubov,Chapman,Ferziger,Resibois,Humenyuk6}.
 However,
 constructing a kinetic theory for ordinary potentials at high densities is significantly complicated
 (there are no small parameters).
 For dense gases and liquids,
 time intervals $\tau_{f}$ --- time of free run and $\tau_{hyd}$ --- characteristic time of change of sizes which are locally stored --- densities mass, momentum, and energy have the same order of magnitude.
 Therefore in the process of ``collision'' of two particles, an important role is played by other system particles.
 This means that the kinetics of dense systems must consider the collective effects that are characteristic of the hydrodynamic approach.
 In other words,
 this means that for dense gases and liquids,
 kinetics and hydrodynamics are closely related and should be considered simultaneously.
 There are only a few model systems for which it is possible to introduce a proper kinetic equation
 (and then in the approximation of paired collisions) and obtain formulas for the transfer coefficients.
 There are only a few model systems for which it is possible to introduce a proper kinetic equation
 (and then in the approximation of paired collisions) and obtain formulas for the transfer coefficients.
 This is a system of solid spheres~\cite{Enskog,Chapman,Ferziger,Beijeren1,Beijeren2,Resibois,Polewczak1,Polewczak2,Polewczak33},
 a system with a square-well potential(SWP)~\cite{Davis,Karkheck,Beijeren3,Wilbertz,Leegwater,Dufty,Garland,Yu,Dwivedee,Takada} and with a multistep potential (MSP)~\cite{Tokarchuk1,Tokarchuk2,Tokarchuk3,omelyan6,kobr3,Humenyuk,Humenyuk1,Humenyuk99} as a generalization of SWP.
 The first model system replaces all the real potential by the effective diameter of solid spheres.
 It is a sure recipe for calculating the contact value of the pair distribution function.
 The latter two are unique in that they explicitly consider the far-reaching part of the potential due to the corresponding irreversible integral of the collisions.

 In other models,
 Enskog's kinetic theory or its revised variant (RET)~\cite{Beijeren1,Beijeren2} is mainly used as a basis,
 and long-range interaction is taken into account indirectly.
 In particular,
 Rice and Allnatt~\cite{Rice} added a Fokker--Planck contribution to the integral of Enskog collisions:
 the particles between collisions on a solid core seem to undergo Brownian motion due to the far-reaching part of the potential.
 Due to this approximation,
 the numerical calculations for shear viscosity and thermal conductivity differ significantly~\cite{Vogelsang} from the data of molecular dynamics.
 Based on the entropy maximization approach with some elms,
 for the potential ``solid spheres + smooth tail'', several kinetic theories of the mean-field type (KTMF) have been proposed~\cite{Karkheck1,Karkheck2,Karkheck3},
 which are parallel and indirect, and take into account the attractive interaction.
 In~\cite{Karkheck4},
 the kinetic theory of the mean-field is applied to mixtures,
 and the result is obtained within the Katz boundary.
 In~\cite{Dyer},
 the calculations of KTMF for the coefficients of self-diffusion,
 shear,
 and bulk viscosity are compared with the results of other indirect approaches --- stochastic kinetic theory and renormalized Kirkwood theory.
 Other works that use semi-phenomenological methods to account for attractive interactions include~\cite{Morioka},
 where an averaged scattering cross-section for viscosity is introduced.
 The work~\cite{Karkheck5} also proposes a kinetic theory for a system of particles with the potential ``hard spheres + truncated tail'',
 which combines a kinetic theory for SWP systems and KTMF.
 This theory gives a good agreement between the numerical calculations for the transfer coefficients and the results of computer simulations and experiments.

 The multistage potential has some advantage over a square-well because it allows us to approximate real interaction better. It should note that in the calculations of the transfer coefficients for argon~\cite{kobr3},
 we have to some extent seen the role of the first coordination sphere in the optimal choice of the number of attraction walls in the multistage potential,
 which simulates the Lennard--Jones potential.
 In fact,
 in these calculations,
 there is a problem of optimal choice of the number of repulsive and attractive walls,
 which is primarily manifested in the pair function of particle distribution,
 the behavior of which determines the first coordination area.
 In fact,
 due to these pairwise particle distribution functions,
 the liquid argon transfer coefficients were calculated.
 It was shown that after the self-consistent optimal number of attraction walls,
 the increase in their number did not significantly affect the calculation results.
 It is obvious that within the first coordination sphere in simple,
 complex liquids,
 there are short-acting repulsive and attractive forces,
 the true nature of which is connected with the internal nuclear-electronic structure of atoms
 and molecules, and has a quantum-mechanical nature (regardless of temperature) in a certain way we model one or another potential of interaction at the classical level of description.
 The kinetic theory of mixtures with a multistage interaction potential is considered in~\cite{Humenyuk99}.
 The boundary transition to a smooth continuous potential for the kinetic equation and the balanced equation for the potential energy density is analyzed.
 When the limit potential is chosen in the form of ``solid spheres + smooth tail'',
 the kinetic equation is reduced to the equation of kinetic variation theory (KVT).
 Still,
 the limit equation of balance for the kinetic energy density differs from the corresponding equation KVT.

 To highlight the contribution of short-range repulsive and attracting particles within the formation of the short order (first coordination sphere) in the theory of nonequilibrium processes,
 we consider a chain of BBGKI equations for nonequilibrium distribution functions of interacting particles of a classical system.
 We use the approach proposed in~\cite{zub5} based on the modification of the boundary conditions for the weakening of correlations to the chain of BBGKI equations.
 This approach was further developed in the works~\cite{84a,zub9,zub10,morozov1,omelyan6,kobr2,tok11,kobr1,kobr3,hlush1,cmp10,cmp12,Markiv}.
 It considers both the nonequilibrium of the one-particle distribution function and the local laws of conservation of mass,
 momentum
 and total energy,
 which form the basis of the hydrodynamic description of the system's evolution.
 Based on the modification of the boundary condition of attenuation of correlations to the Liouville equation,
 a system of coupled generalized equations for the nonequilibrium one-particle distribution function and the average total energy density was obtained,
 which are valid for describing nonequilibrium states of both far and near equilibrium.
 An important achievement of this approach is that the formulated modified boundary condition,
 which takes into account local conservation laws to the chain BBGKI,
 made it possible for the first time in~\cite{zub9} to consistently derive the kinetic equation of the revised Enskog theory~\cite{Beijeren1,Beijeren2,Resibois} for the system of solid spheres.
 Based on this result,
 we obtained a kinetic equation for a system with a model multistep interparticle interaction potential~\cite{omelyan6}
 (in particular,
 for this kinetic equation was proved $H$-theorem in~\cite{omelyan6}),
 as well as the kinetic equation Landau~\cite{84a,zub9,zmot} for a system of charged solid spheres.
 For the new kinetic equations obtained in this way,
 normal solutions were found by the Chapman--Enskog method,
 which was used to perform numerical calculations of the transfer coefficients of bulk,
 shear viscosities,
 and thermal conductivity for systems that simulate argon~\cite{zmot,kobr3},
 once ionized argon~\cite{kobr2}.

 An important problem in applying the obtained kinetic equations with integrals of collisions in approximations of pair collisions
 or polarization is the calculation of the pair quasi-equilibrium distribution function $g_2(\vec{r}_1,\vec{r}_2|n,\beta;t)$ as functions of nonequilibrium values:
 of average density $n(\vec{r};t)$ and inverse temperature $\beta(\vec{r};t)$.
 In the works~\cite{Karkheck3,Polewczak1,Polewczak33},
 when considering the corresponding models of the collision integral for this distribution function,
 we used the generalization of the virial decomposition by the density chosen over time.

 In the second section,
 we consider a chain of BBGKI equations with a modified boundary condition and with the selection of short-range repulsive and attractive interactions between particles of the classical system.
 In the third section,
 we consider a chain of BBGKI equations with a modified boundary condition to approximate ``paired '' collisions and obtain the corresponding kinetic equation.
 In this section,
 we obtain the kinetic equation for the nonequilibrium one-particle distribution function with the collision integral in the polarization approximation when the interaction of two particles is taken into account through the interaction of the third particle.
 In the fourth section,
 we present one of the ways to calculate the pair quasi-equilibrium distribution function,
 which is included in the collision integrals of the corresponding kinetic equations.

 \vspace{-1mm}
 \section{Chain of BBGKI equations for nonequilibrium particle distribution functions of a simple liquid}
 \vspace{-1mm}

 As a model of a simple fluid,
 we will consider a system of classical interacting particles with the following Hamiltonian:
 \begin{equation*} %      \label{e.01}
  H=\sum_{j=1}^{N}\frac{\vec{p}_{j}}{2m} +\frac{1}{2}\sum_{j,l}^{N}\Phi(\vec{r}_{j},\vec{r}_{l})
 \end{equation*}\\[-3mm]
 where the indices $j$, $l$ number the particles with mass $m$ with vector pulses $\vec{p}_{j}$ and vectors with coordinates $\vec{r}_{j}$ simple liquid,
 $N$ is the total number of particles in the volume $V$;\vspace{-2mm}
 \begin{equation}       \label{e.0102}
  \Phi(\vec{r}_{j},\vec{r}_{l})=\Phi_{rep}(\vec{r}_{j},\vec{r}_{l})+\Phi_{attr}(\vec{r}_{j},\vec{r}_{l})
  =
  \Phi_{rep}^{sr}(\vec{r}_{j},\vec{r}_{l})+\Phi_{attr}^{sr}(\vec{r}_{j},\vec{r}_{l})+\Phi_{attr}^{lr}(\vec{r}_{j},\vec{r}_{l}),
 \end{equation}\\[-7mm]
 there is a pair potential for interaction between particles,
 consisting of a short-acting repulsive
 (for example, a model of solid spheres) $\Phi_{rep}^{sr}(\vec{r}_{j},\vec{r}_{l})$,
 short-range attractive $\Phi_{attr}^{sr}(\vec{r}_{j},\vec{r}_{l})$ and long-range attractive parts $\Phi_{attr}^{lr}(\vec{r}_{j},\vec{r}_{l})$.
 The full short-term interaction is equal to\vspace{-2mm}
 \[
  \Phi_{tot}^{sr}(\vec{r}_{j},\vec{r}_{l})=\Phi_{rep}^{sr}(\vec{r}_{j},\vec{r}_{l})+\Phi_{attr}^{sr}(\vec{r}_{j},\vec{r}_{l}).
 \]\\[-12mm]

 The nonequilibrium state of the system is completely described by the nonequilibrium distribution function of all particles
 $\rho(x_{1},\ldots,x_{N};t)=\rho(x^{N};t)$
 (where the notation $x^{N}=x_{1},\ldots,x_{N}$,
 $x_{j}=\vec{p}_{j}\vec{r}_{j}$),
 whose evolution in time is given Liouville equation\vspace{-2mm}
 \begin{equation*}%\label{e.001}
  \frac{\partial }{\partial t}\rho(x^{N};t)+iL_{N}\rho(x^{N};t)=0.
 \end{equation*}\\[-6mm]
 with the Liouville operator\vspace{-2mm}
 \begin{equation*} %\label{e.02}
  iL_{N}=\sum_{j=1}^{N}\frac{\vec{p}_{j}}{m}\cdot\frac{\partial }{\partial \vec{r}_{j}}
        -\sum_{j,l}^{N}\frac{\partial }{\partial \vec{r}_{j}}\Phi(\vec{r}_{j},\vec{r}_{l})
          \left(\frac{\partial }{\partial \vec{p}_{j}}- \frac{\partial }{\partial \vec{p}_{l}}\right) .
 \end{equation*}\\[-4mm]
 We will describe the nonequilibrium state of the system with the help of chain of BBGKI equations for nonequilibrium fluid particle distribution functions.
 To do this,
 we use the approach proposed in~\cite{zub9,tok11},
 where a modified chain of equations BBGKI is built,
 taking into account the concept of a consistent description of kinetics and hydrodynamics of nonequilibrium processes of the system of interacting particles in the method of nonequilibrium statistical operator Zubarev,
 based on the Liouville equation with the source:\vspace{-2mm}
 \begin{equation}\label{e.0010}
  \frac{\partial}{\partial t}\rho(x^{N};t)+iL_{N}\rho(x^{N};t)=-\varepsilon(\rho(x^{N};t)-\varrho_{rel}(x^{N};t)),
 \end{equation}\\[-5mm]
 which selects the delayed ($\varepsilon\rightarrow +0$,
 after thermodynamic transition) solutions of the Liouville equation under given initial conditions:
 \[
  \rho(x^{N};t)_{t=t_{0}}=\varrho_{rel}(x^{N};t_{0}),
 \]\\[-6mm]
 $\rho_{rel}(x^{N_{\alpha}};t)$ is the relevant (quasi-equilibrium) particle distribution function obtained according to~\cite{zub9,tok11} from the condition of the maximum of the Gibbs entropy functional with preserved normalization conditions for the distribution and the set parameters of the abbreviated description of the nonequilibrium state of the liquid:
 $\langle\hat{n}(x)\rangle^{t}=f_{1}(x;t)$ is the nonequilibrium one-particle particle distribution function of liquid and $\langle \hat{\varepsilon}_{int}(\vec{r})\rangle^{t}$ is the nonequilibrium average energy of the interaction of liquid particles,
 which has the following structure:\vspace{-2mm}
 \begin{equation}       \label{e.002}
  \rho_{rel}(x^{N};t)=\exp\left\{-\Phi(t)-\int d\vec{r}\,\beta(\vec{r},t)\hat{\varepsilon}_{int}(\vec{r})
  -\int dx \,a(x,t)\hat{n}(x)\right\},
 \end{equation}\\[-5mm]
 where $\Phi(t)$ is the Masier--Planck functional\vspace{-2mm}
 \begin{align*}      % \label{e.003}
  \Phi(t)&=\int d\Gamma_{N}(x) \exp\left\{-\int d\vec{r}\,\beta(\vec{r},t)\hat{\varepsilon}_{int}(\vec{r})
  -\int dx \,a(x,t)\hat{n}(x)\right\}
  =\ln Z_{rel}(t) ,\\  %\label{e.003a}
  Z_{rel}(t)&=\int d\Gamma_{N}(x)
  \exp\left\{-\int d\vec{r}\,\beta(\vec{r},t)\hat{\varepsilon}_{int}(\vec{r})
  -\int dx \,a(x,t)\hat{n}(x)\right\}
 \end{align*}\\[-5mm]
 is the statistical sum of the quasi-equilibrium distribution $\rho_{rel}(x^{N};t)$,
 in which\vspace{-3mm}
 \begin{equation*}   %    \label{e.004}
  \hat{n}(x)=\sum_{j=1}^{N}\delta (x-x_{j})=\sum_{j=1}^{N}\delta (\vec{r}-\vec{r}_{j})\delta (\vec{p}-\vec{p}_{j})
 \end{equation*}\\[-5mm]
 is the microscopic phase density of the number of particles and\vspace{-3mm}
 \begin{equation*}   %    \label{e.005}
  \hat{\varepsilon}_{int}(\vec{r})=\frac{1}{2}\sum_{j,l}^{N}\Phi(\vec{r}_{j},\vec{r}_{l}) \delta (\vec{r}-\vec{r}_{j})
  =\hat{\varepsilon}_{rep}^{sr}(\vec{r})+ \hat{\varepsilon}_{attr}^{sr}(\vec{r})+\hat{\varepsilon}_{attr}^{lr}(\vec{r})
 \end{equation*}\\[-5mm]
 is the microscopic energy density of the interaction of particles with the selected contributions of repulsive and attractive interactions;
 $d\Gamma=(dx)^N/N!$,
 $dx=d\vec{p}\,d\vec{r}$.
 Moreover,
 the sum of the first two $\hat{\varepsilon}_{tot}^{sr}(\vec{r})$ form the microscopic energy density of the interaction of particles in the first coordination sphere.
 The Lagrange parameters $\beta(\vec{r},t)$ (inverse of the nonequilibrium temperature of the liquid),
 $a(x,t)$ are determined from the conditions of self-agreement:\vspace{-2mm}
 \begin{equation}       \label{e.0050}
  \langle \hat{\varepsilon}_{int}(\vec{r})\rangle^{t}=\langle \hat{\varepsilon}_{int}(\vec{r})\rangle^{t}_{rel},\qquad    \langle\hat{n}(x)\rangle^{t}= \langle\hat{n}(x)\rangle^{t}_{rel}.
 \end{equation}\\[-7mm]
 Taking into account,
 the structure $\rho_{rel}(x^{N};t)$ and the approach~\cite{zub9,tok11},
 integrating the Liouville equation with the source~(\ref{e.0010}) by the corresponding coordinates and momentum of the particles,
 we obtain chain of BBGKI equations with modified boundary conditions
 (taking into account space-time interparticle correlations) for a simple fluid:\vspace{-2mm}
 \begin{align} \label{e.03}
  &\left(\frac{\partial }{\partial t} +iL(1)\right)f_{1}(x_{1};t)
   +\int dx_{2}\,iL(1,2) f_{2}(x_{1},x_{2};t)=0,\\ \label{e.04}
  &\left(\frac{\partial }{\partial t}+iL(1)+iL(2)+iL(1,2)\right)f_{2}(x_{1},x_{2};t)
    +\int dx_{3} \big(iL(1,3) +iL(2,3)\big)
    f_{3}(x_{1},x_{2},x_{3};t) \nonumber\\
  &\qquad\qquad\qquad\qquad\qquad\qquad\qquad\qquad\quad
    =- \varepsilon \big(f_{2}(x_{1},x_{2};t)-g_{2}(\vec{r}_{1},\vec{r}_{2}|n,\beta ;t)
 f_{1}(x_{1};t)f_{2}(x_{2};t)\big),
 \end{align}\\[-7mm]
 where $\varepsilon \rightarrow +0$ after the limit thermodynamic transition,\vspace{-2mm}
 \begin{align*}% \label{e.05}
   iL(j)  &= \frac{\vec{p}_{j}}{m}\cdot\frac{\partial }{\partial \vec{r}_{j}}, \nonumber\\
   iL(j,l)&= -\frac{\partial }{\partial  \vec{r}_{j}}\Phi(\vec{r}_{j},\vec{r}_{l})\cdot
              \left(\frac{\partial }{\partial \vec{p}_{j}}- \frac{\partial }{\partial \vec{p}_{l}} \right)\nonumber\\
          &= -\frac{\partial }{\partial\vec{r}_{j}}\left(\Phi_{rep}^{sr}(\vec{r}_{j},\vec{r}_{l})
             +\Phi_{attr}^{sr}(\vec{r}_{j},\vec{r}_{l})+\Phi_{attr}^{lr}(\vec{r}_{j},\vec{r}_{l})\right)
             \cdot\left(\frac{\partial }{\partial \vec{p}_{j}}- \frac{\partial }{\partial \vec{p}_{l}} \right),
 \end{align*}\\[-5mm]
 there are one-part and two-part parts of the Liouville operator,
 $g_{2}(\vec{r}_{1}, \vec{r}_{2}|n;t)$ is a pair of quasi-equilibrium coordinate functions of the distribution of fluid particles\vspace{-2mm}
 \begin{equation} \label{e.0606}
  g_{2}(\vec{r}_{1}, \vec{r}_{2}|n,\beta ;t)= \frac{1}{n(\vec{r}_{1};t)n(\vec{r}_{2};t)}\int d\Gamma_{N}(x)\,\hat{n}(\vec{r}_{1})\,\hat{n}(\vec{r}_{2})\,\rho_{rel}(x^{N};t).
 \end{equation}\\[-5mm]
 The distribution functions of the three particles satisfy the following equations of the chain of equations BBGKI,
 which include nonequilibrium functions of the distribution of four particles.
 It is important to note that one- and two-particle nonequilibrium distribution functions determine the behavior of hydrodynamic variables:
 average nonequilibrium values densities of the number of particles $n(\vec{r};t)$,
 their momentum $\vec{p}(\vec{r};t)$,
 kinetic energy $\varepsilon^{kin}(\vec{r};t)$,
 as well as the potential energy $\varepsilon^{int}(\vec{r};t)$:\vspace{-2mm}
 \begin{align*} %\label{e.070}
  n(\vec{r};t)&=\int d\vec{p}\,f_{1}(\vec{r},\vec{p};t), \qquad
  \vec{p}(\vec{r};t)=\int d\vec{p}\,f_{1}(\vec{r},\vec{p};t)\vec{p},\\ %\label{e.071}
  \varepsilon^{kin}(\vec{r};t)&=\int d\vec{p}\,
  \frac{p^{2}}{2m}f_{1}(\vec{r},\vec{p};t), \\ \nonumber
  \varepsilon^{int}(\vec{r};t)&=\int d\vec{p}\int d\vec{p}'
  \int d\vec{r}'\,\Phi(\vec{r},\vec{r}')
  f_{2}(\vec{r},\vec{p},\vec{r}',\vec{p}';t),
 \end{align*}\\[-5mm]
 which satisfy the corresponding conservation laws of the average nonequilibrium values of the number of particles of a simple liquid $n(\vec{r};t)$,
 their momentum $\vec{p}(\vec{r};t)$ and full of energy\vspace{-2mm}
 \begin{equation*}% \label{e.073}
  \varepsilon(\vec{r};t)=\varepsilon^{kin}(\vec{r};t)+\varepsilon^{int}(\vec{r};t),
 \end{equation*}\\[-7mm]
 underlying the hydrodynamic description of nonequilibrium processes of a simple fluid.
 In addition,
 the pair is a quasi-equilibrium coordinate distribution function $g_{2}(\vec{r},\vec{r}'|n,\beta;t)$ and higher coordinate quasi-equilibrium distribution functions $g_{3}(\vec{r},\vec{r}',\vec{r}''| n,\beta;t)$, \ldots,
 $g_{s}(\vec {r}^{s}| n,\beta;t)$ in the following equations BBGKI chains describe multiparticle correlations and are functions of nonequilibrium densities $n(\vec{r};t)$ and temperature $\beta(\vec{r};t)$.
 The structure of the data of correlation functions should reflect the features of the selected model for the pair potential of particle interaction~(\ref{e.0102}).
 One of the ways to calculate them will be considered in the last section.

 In the next section,
 we consider the chain of BBGKI equations in the approximation of pairwise collisions between particles.

 \section{The chain of BBGKI equations in the approximation of pairwise collisions of particles and in the polarization approximation}

 In the approximation of pairwise collisions between particles,
 when the distribution functions of the three particles are not taken into account,
 for nonequilibrium two-particle distribution functions,
 we obtain~\cite{zub9,tok11} the following equation:\vspace{-2mm}
 \begin{equation} \label{e.07}
  \bigg(\frac{\partial }{\partial t} +iL(1)+iL(2)+iL(1,2)\!\bigg)\!
   f_{2}(x_{1},x_{2};t)
  =- \varepsilon \big(f_{2}(x_{1},x_{2};t)-
   g_{2}(\vec{r}_{1}, \vec{r}_{2}|n,\beta ;t)f_{1}(x_{1};t)f_{1}(x_{2};t)\big).
 \end{equation}\\[-4mm]
 The solution of this equation (\ref{e.07}) can be represented as\vspace{-2mm}
 \begin{equation*}% \label{e.09}
  f_{2}(x_{1},x_{2};t)=\varepsilon \int_{-\infty}^{0} d\tau\,
  e^{(\varepsilon +i\bar{L}(1,2))\tau}
  g_{2}(\vec{r}_{1},\vec{r}_{2}|n,\beta ;t+\tau)
  f_{1}(x_{1};t+\tau)f_{1}(x_{2};t+\tau),
 \end{equation*}\\[-4mm]
 where
 \begin{equation*}
  i\bar{L}(1,2)= iL(1)+iL(2)+iL(1,2).
 \end{equation*}\\[-5mm]
 Substituting this solution into the equation~(\ref{e.03}),
 we obtain a non-Markov kinetic equation for nonequilibrium one-particle simple fluid distribution function:\vspace{-2mm}
 \begin{equation*}% \label{e.011}
  \left(\frac{\partial }{\partial t} +iL(1)\!\right)\!f_{1}(x_{1};t)=
   -\int \!\!dx_{2}\,iL(1,2) \varepsilon
    \int_{-\infty}^{0}\!\!\!\!
   d\tau \,e^{(\varepsilon +i\bar{L}(1,2))\tau}
   g_{2}(\vec{r}_{1}, \vec{r}_{2}|n,\beta ;t+\tau)
   f_{1}(x_{1};t+\tau)f_{1}(x_{2};t+\tau).
 \end{equation*}\\[-4mm]
 The obtained kinetic equation for the nonequilibrium one-particle particle distribution function is taken into account
 also the spatial heterogeneity of the system.
 According to the nature of the interaction of particles~(\ref{e.0102}),
 the integral of the collisions in the right part is given in the form:\vspace{-2mm}
\begin{align} \label{e.0110}
 I(x;t)&=-\int dx_{2}\, iL(1,2) \varepsilon \int_{-\infty}^{0} d\tau \,e^{(\varepsilon +i\bar{L}(1,2))\tau}
         g_{2}(\vec{r}_{1}, \vec{r}_{2}|n,\beta ;t+\tau) f_{1}(x_{1};t+\tau)f_{1}(x_{2};t+\tau) \\
       &=-\int d\vec{p}_{2}\bigg(\int_{0}^{\sigma}iL_{rep}^{sr}(1,2)\,\varepsilon \int_{-\infty}^{0}
          d\tau \,e^{(\varepsilon +i\bar{L}_{rep}^{sr}(1,2))\tau}
        +\int_{\sigma}^{r_{min}}iL_{attr}^{sr}(1,2)\, \varepsilon \int_{-\infty}^{0}
  d\tau e^{(\varepsilon +i\bar{L}_{attr}^{sr}(1,2))\tau}\nonumber \\
 &+\int_{r_{min}}^{\infty}iL_{attr}^{lr}(1,2)\, \varepsilon \int_{-\infty}^{0}
  d\tau \,e^{(\varepsilon +i\bar{L}_{attr}^{lr}(1,2))\tau}\bigg)
  g_{2}(\vec{r}_{1}, \vec{r}_{2}|n\beta ;t+\tau) f_{1}(x_{1};t+\tau)f_{1}(x_{2};t+\tau)\,d\vec{r}_{2},\nonumber
 \end{align}\\[-4mm]
 where the first term of the collision integral describes the repulsive short-range processes of particle scattering,
 and the second represents the attractive short-range scattering processes in the approximate pair interactions.
 They form the kinetics of the middle order (the first coordination sphere) in a simple fluid.
 The third term in the collision integral describes the attractive long-range processes of particle scattering.
 It forms a long-range order with a consistent description of the kinetics and hydrodynamics of a simple fluid.
 In the future,
 we will choose the repulsive short-range interaction potential in the form of a model of solid spheres:\vspace{-2mm}
 \begin{equation*}%\label{2.2.15}
  \Phi_{rep}^{sr}(|\vec{r}_{jl}|)=\Phi^{h.s.}(|\vec{r}_{jl}|)
  =
  \lim_{\epsilon\rightarrow \infty}\Phi^{\epsilon}(|\vec{r}_{jl}|),
 \end{equation*}\\[-6mm]
 where
 \begin{equation*}%\label{2.2.16}
  \Phi^{\epsilon}(|\vec{r}_{jl}|)
  =
  \left\{%
  \begin{array}{ll}
     \epsilon, & |\vec{r}_{jl}|<\sigma; \\
    0, & |\vec{r}_{jl}|\geq\sigma. \\
  \end{array}%
  \right.
 \end{equation*}
 $\sigma$ is the diameter of a solid sphere.
 The potential $\Phi^{h.s.}(|\vec{r}_{jl}|)$ for solid spheres is strongly singular
 (in particular, the operator Liouville $iL(j,l)$ is poorly defined).
 That's why we will consider the potential $\Phi^{\epsilon}(|\vec{r}_{jl}|)$ and only in the final expressions we put $\lim_{\epsilon \rightarrow\infty}$.
 Note the characteristics of the potential $\Phi^{h.s.}(|\vec{r}_{jl}|)$,
 $\Phi^{\epsilon}(|\vec{r}_{jl}|)$.
 First,
 since the radius of their interaction is $r_{0}=\sigma^{+}$,
 then the area of interaction $\Delta r_{0}\rightarrow +0$,
 and,
 secondly,
 time interaction $\tau_{0}\rightarrow +0$, which is due to the singular the nature of the potential.
 Then,
 according to the approach~\cite{zub9,tok11},
 this part of the collision integral can be represented in the form of the collision integral in the revised theory of Enskog:\vspace{-2mm}
 \begin{align} \label{e.0111}
  I_{rep}^{sr}(x;t)&=-\!\!\int \!\!d\vec{p}_{2}\int_{0}^{\sigma}\!\!iL_{rep}^{sr}(1,2)\,\varepsilon \int_{-\infty}^{0}\!\!\!\!\! d\tau e^{(\varepsilon +i\bar{L}_{rep}^{sr}(1,2))\tau} g_{2}(\vec{r}_{1}, \vec{r}_{2}|n,\beta ;t+\tau)f_{1}(x_{1};t+\tau)f_{1}(x_{2};t+\tau)d\vec{r}_{2}\nonumber\\
  &=-n\int dx_{2}\,{\hat T}(1,2)g_{2}(\vec{r}_{1},\vec{r}_{2}|n,\beta ;t)f_{1}(x_{l};t)f_{1}(x_{2};t),
 \end{align}\\[-7mm]
 where
 \begin{equation*}%\label{2.2.66}
  {\hat T}(1,2)=\sigma^{2}\int d \hat{\vec{\sigma}}\,\Theta(\hat{\vec{\sigma}}\cdot \vec{g}) (\hat{\vec{\sigma}}\cdot \vec{g})\left[\delta(\vec{r}_{1}-\vec{r}_{2}+ \sigma^{+}\hat{\vec{\sigma}}){\hat
   B}(\vec{\sigma})- \delta(\vec{r}_{1}+\vec{r}_{2}- \sigma^{+}\hat{\vec{\sigma}})\right]
 \end{equation*}\\[-5mm]
 is the operator of collisions of solid elastic spheres of Enskog,
 where\vspace{-3mm}
 \begin{equation}\label{2.2.64}
   \vec{v}'_{1}=\vec{v}_{1}+ \hat{\vec{\sigma}}(\hat{\vec{\sigma}}\cdot \vec{g}),\qquad
   \vec{v}'_{2}=\vec{v}_{2}- \hat{\vec{\sigma}}(\hat{\vec{\sigma}}\cdot \vec{g}),
 \end{equation}\\[-8mm]
 $\vec{v}_{1}$, $\vec{v}_{2}$ are vectors --- the velocities of particles to the interaction,
 and $\vec{v}'_{1}$, $\vec{v}'_{2}$ are vectors --- their velocities after interaction with the relative velocity
 $\vec{g}= \vec{v}_{2}-\vec{v}_{1}$,
 ${\hat{B}}(\hat{\sigma})\varphi(\vec{v}_{1},\vec{v}_{2})=\varphi(\vec{v}'_{1},\vec{v}'_{2})$
 is the speed shift operator defined ratios~(\ref{2.2.64}).
 Further,
 if in the second and third terms of the collision integral~(\ref{e.0110}) perform the integration in parts and take into account~(\ref{e.0111}),
 the collision integral can be represented as:\vspace{-3mm}
 \begin{align} \label{e.0112}
  I(x;t)&=-n\int dx_{2}\,{\hat T}(1,2)\,g_{2}(\vec{r}_{1},\vec{r}_{2}|n,\beta ;t)\,f_{1}(x_{l};t)\,f_{1}(x_{2};t) \nonumber\\
  &\quad-\int d\vec{p}_{2}\int_{\sigma}^{r_{\min}}iL_{attr}^{sr}(1,2)\,g_{2}(\vec{r}_{1},\vec{r}_{2}|n;t)\,f_{1}(x_{1};t)\,f_{1}(x_{2};t) d\vec{r}_{2} \nonumber \\
  &\quad+\int d\vec{p}_{2}\int_{\sigma}^{r_{\min}}iL_{attr}^{sr}(1,2)
  \int_{-\infty}^{0}d\tau \,e^{(\varepsilon +i\bar{L}_{attr}^{sr}(1,2))\tau}
  \left(\frac{\partial }{\partial t}+i\bar{L}_{attr}^{sr}(1,2)\right)\nonumber \\
  &\qquad\qquad\times g_{2}(\vec{r}_{1}, \vec{r}_{2}|n,\beta ;t+\tau) f_{1}(x_{1};t+\tau)f_{1}(x_{2};t+\tau) \,d\vec{r}_{2}\nonumber \\
  &\quad-\int d\vec{p}_{2}\int_{r_{\min}}^{\infty}iL_{attr}^{lr}(1,2)\,g_{2}(\vec{r}_{1},\vec{r}_{2}|n;t)\,f_{1}(x_{1};t)\,f_{1}(x_{2};t)\, d\vec{r}_{2}\nonumber \\
  &\quad+\int d\vec{p}_{2}\int_{r_{\min}}^{\infty}iL_{attr}^{lr}(1,2)  \int_{-\infty}^{0}
  d\tau\, e^{(\varepsilon +i\bar{L}_{attr}^{lr}(1,2))\tau}\left(\frac{\partial }{\partial t}+i\bar{L}_{attr}^{lr}(1,2)\right)\nonumber \\
  &\qquad\qquad\times  g_{2}(\vec{r}_{1}, \vec{r}_{2}|n,\beta ;t+\tau)\,f_{1}(x_{1};t+\tau)\,f_{1}(x_{2};t+\tau)\, d\vec{r}_{2},
 \end{align}\\[-8mm]
 in which the second and fourth terms in the right part correspond to the generalized middle fields with potentials of short-range and long-range attraction.
 The third and fifth terms of the collision integral are of the second order in terms of interactions taking into account non-Markov processes (time delay).

 The kinetic equation with the collision integral~(\ref{e.0112}) is derived from the BBGKI chain of equations in approaching ``pair'' collisions,
 although a significant part of the coordinates of dynamic correlations is already taken into account in the quasi-equilibrium distribution functions $g_2(\vec{r}_1,\vec{r}_2|n,\beta ;t)$.

 For analysis of chain connections of BBGKI equations (\ref{e.03})--(\ref{e.04}) in higher approximations by interparticle correlations it is convenient to apply the concept of group compositions~\cite{G3,G4,G10}.
 Group deposits were applied to the chain of equations of BBGKI in many works~\cite{G3,G4,G5,G6,G7,G8,Vl,G9,G10,G12} with a boundary condition corresponding to the attenuation condition correlations by Bogolyubov,
 in the works of Zubarev and Novikov~\cite{G10,G11,G111},
 which he was developed diagram method for constructing connections of the chain of BBGKI equations.
 For the connection of the chain of equations (\ref{e.03})--(\ref{e.04}) similarly as in the works~\cite{G10,G11,G111},
 and even more so in the works~\cite{G3,G4,G5},
 let's move from nonequilibrium of distribution functions $f_{s}(x^{s};t)$ to irreducible distribution functions
 $F_{s} (x^{s};t)$,
 which are introduced by the quantities~\cite{G3,G10},
 but in in our case with some modification:\vspace{-2mm}
 \begin{align}\label{e4.1}
  f_1(x_1;t)&=F_1(x_1;t), \\
  f_2(x_1,x_2;t)&=F_2(x_1,x_2;t)+   g_2(\vec{r}_1,\vec{r}_2;t)F_1(x_1;t)F_1(x_2;t),    \nonumber\\
  f_3(x_1,x_2,x_3;t)&=F_3(x_1,x_2,x_3;t) \nonumber\\
                    &\quad+\sum\nolimits_{P}F_2(x_1,x_2;t)F_1(x_3;t)     +g_3(\vec{r}_1,\vec{r}_2,\vec{r}_3;t)F_1(x_1;t)F_1(x_2;t)F_1(x_3;t),\quad  \ldots ,  \nonumber
 \end{align}\\[-6mm]
 in which the coordinate quasi-equilibrium distribution functions
 $g_2(\vec{r}_1,\vec{r}_2;t)$,
 $g_3(\vec{r}_1,\vec{r}_2,\vec{r}_3;t)$,
 $g_s(\vec{r}^s;t)$ are determined by the relations~(\ref{e.0606}) and\vspace{-2mm}
 \begin{equation*}% \label{e.0606a}
  g_{s}(\vec{r}^{s}|n,\beta;t)= \frac{1}{n(\vec{r}_{1};t)\ldots n(\vec{r}_{s};t)}\int d\Gamma_{N}(x)\hat{n}(\vec{r}_{1})\ldots\hat{n}(\vec{r}_{s})\rho_{rel}(x^{N};t).
 \end{equation*}\\[-9mm]

 The modification of group compositions (\ref{e4.1}) is that a significant part of simple correlations is taken into account over time quasi-equilibrium functions $g_s(\vec{r}^s;t)$.
 When $g_s(\vec{r}^s;t)=1$ for $s=2,3,\ldots$ these group clauses coincide with group clauses~\cite{G3,G4,G5}.
 Since each line of equivalences (\ref{e4.1}) introduces exactly one new one function $F_s(\vec{r}^s;t)$,
 $s=1,2,3,\ldots$,
 then the data of the equation can be related relatively of irreducible separation functions and write:\vspace{-3mm}
 \begin{align}\label{e4.2}
  F_1(x_1;t)&=f_1(x_1;t), \\
  F_2(x_1,x_2;t)&=f_2(x_1,x_2;t)-g_2(\vec{r}_1,\vec{r}_2;t)f_1(x_1;t)f_1(x_2;t),    \nonumber\\
  F_3(x_1,x_2,x_3;t)&=f_3(x_1,x_2,x_3;t)-\sum\nolimits_{P}f_2(x_1,x_2;t)f_1(x_3;t)    \nonumber\\
                    &\quad-h_3(\vec{r}_1,\vec{r}_2,\vec{r}_3;t)f_1(x_1;t)f_1(x_2;t)f_1(x_3;t), \quad  \ldots .
 \end{align}\\[-7mm]
 Here and in (\ref{e4.1}) $\sum_P$ denotes the sum of all different permutations the coordinate of three or more particles,\vspace{-3mm}
 \begin{equation}\label{e4.3}
  h_3(\vec{r}_1,\vec{r}_2,\vec{r}_3;t)=g_3(\vec{r}_1,\vec{r}_2,\vec{r}_3;t)-g_2(\vec{r}_1,\vec{r}_2;t)-
    g_2(\vec{r}_1,\vec{r}_3;t)-g_2(\vec{r}_2,\vec{r}_3;t)
 \end{equation}\\[-7mm]
 is a three-particle quasi-equilibrium correlation function.
 Now we write down the chain of equations BBGKI (\ref{e.03})--(\ref{e.04}) for irreducible distribution functions $F_s(x^s;t)$,
 the first two comparison.
 The first equation of the chain has the form:\vspace{-2mm}
 \begin{multline} \label{e4.4}
  \left(\frac{\partial }{\partial t}+iL(1)\right) F_1(x_1;t)
  +\int dx_2\,iL(1,2)\,g_2(\vec{r}_1,\vec{r}_2;t)\,F_1(x_1;t)\,F_1(x_2;t)\\
  +\int dx_2\,iL(1,2)\,F_2(x_1,x_2;t)=0.
 \end{multline}\\[-5mm]
 Differentiating by time the expression for $F_2(x_1,x_2;t)$ in (\ref{e4.2}) and using the second equation of the BBGKI chain~(\ref{e.04}) for the function $f_2(x_1,x_2;t)$, for $F_2(x_1,x_2;t)$,
 the following equation:\vspace{-2mm}
 \begin{multline}\label{e4.5}
  \left(\frac{\partial }{\partial t}+iL_2+\varepsilon\right)
  F_2(x_1,x_2;t)=
  -\left(\frac{\partial }{\partial t}+iL_2\right) g_2(\vec{r}_1,\vec{r}_2;t)F_1(x_1;t)F_1(x_2;t)    \\
  -\int dx_3\Big\{iL(1,3)+iL(2,3)\Big\}\Big\{F_3(x_1,x_2,x_3;t)
  +\sum\nolimits_{P} F_2(x_1,x_2;t)F_1(x_3;t)\\
  +g_3(\vec{r}_1,\vec{r}_2,\vec{r}_3;t)F_1(x_1;t)F_1(x_2;t)F_1(x_3;t)\Big\}.
 \end{multline}\\[-6mm]
 In a similar way,
 you can get an equation for a three-particle of the irreducible function $F_3(x_1, x_2, x_3;t)$ and higher functions $F_s(x^s;t)$ of particle distribution.
 Recall that the emergence of quasi-equilibrium functions distribution $g_2(\vec{r}_1,\vec{r}_2;t)$,
 $g_3(\vec{r}_1,\vec{r}_2, \vec{r}_3;t)$ and $g_s(\vec{r}^s; t)$ in the chain of equations~(\ref{e4.4}),
 (\ref{e4.5}) due to the fact that the boundary condition for the binding of the equation Liuville~(\ref{e.0010}) is considered an imbalance one-particle distribution functions and local security laws,
 which corresponds to the agreed description of kinetics and hydrodynamics systems~\cite{zub5,84a,zub9,zub10,tok11,cmp10,cmp12}.
 Because in this work we will analyze only the first two equations~(\ref{e4.4}), (\ref{e4.5}),
 then the following we will not write down the chain.
 It is necessary note that if put formally $g_s(\vec{r}^s;t)=1$,
 $s = 2,3,\ldots$ in (\ref{e4.4}), (\ref{e4.5}),
 then we get the first two equations of the BBGKI chain of equations for irreducible distribution functions $F_1(x_1;t)$,
 $F_2(x_1,x_2;t)$,
 taken in the work~\cite{G11,G111}.
 A characteristic feature systems of equations~(\ref{e4.4}),
 (\ref{e4.5}) is the first term in the right part of the equation (\ref{e4.5}),
 ie the term with the time derivative from the double quasi-equilibrium distribution function $g_2(\vec{r}_1, \vec{r}_2;t)$. As we already are marked according to (\ref{e.0606}),
 the quasi-equilibrium pair the distribution function is a functional of local values temperatures $\beta(\vec{r};t)$ and average particle density $n(\vec{r};t)$.
 Therefore,
 derivatives in time from $g_2\big(\vec{r}_1, \vec{r}_2|\beta(t), n(t)\big)$ will apply to $\beta(\vec{r};t)$ and $n(\vec{r};t)$,
 which in turn under the terms of self-agreement~(\ref{e.0050}) will be to be expressed through the average values of energy in the monitored system counting $\langle \hat{\cal E}'(\vec{r})\rangle^t$ and $\langle\hat{n}(\vec{r})\rangle^t$,
 which form the basis of the hydrodynamic description nonequilibrium state of the system.

 Now consider the system of equations (\ref{e4.4}), (\ref{e4.5}) in approximation,
 when the second equation~(\ref{e4.5}) does not take into account three-particle irreducible distribution functions
 $F_3(x_1,x_2,x_3;t)$,
 $h_3(\vec{r}_1,\vec{r}_2,\vec{r}_3;t)$,
 which is analogous polarization approximation in the case of retrieval of the Bogolyubov--Lenard--Balescu kinetic equation~\cite{Bogolubov,G32,G61,G62} for Coulomb plasma in the same case.
 Taking into account~(\ref{e4.3}),
 and that $F_3(x_1,x_2,x_3;t)\equiv 0$,
 and also $h_3(\vec{r}_1,\vec{r}_2,\vec{r}_3;t)\equiv 0$.
 The equation~(\ref{e4.5}) is written as follows:\vspace{-3mm}
 \begin{multline}\label{e4.9}
  \left(\frac{\partial }{\partial t}+iL_2+\varepsilon\right)F_2(x_1,x_2;t)
  =
  \left(\frac{\partial }{\partial t}+iL_2\right)
    g_2(\vec{r}_1,\vec{r}_2|n,\beta ;t)F_1(x_1;t)F_1(x_2;t) \\
  -\int dx_3 \, iL(1,3)\left\{F_2(x_1,x_2;t)F_1(x_3;t)+F_2(x_2,x_3;t)F_1(x_1;t)\right\} \\
  -\int dx_3 \, iL(2,3)\left\{F_2(x_1,x_2;t)F_1(x_3;t)+F_2(x_1,x_3;t)F_1(x_2;t)\right\} \\
  -\int dx_3 \left\{iL(1,3)+iL(2,3)\right\}\left\{g_2(\vec{r}_1,\vec{r}_2|n,\beta ;t)+
    g_2(\vec{r}_1,\vec{r}_3|n,\beta ;t)+
    g_2(\vec{r}_2,\vec{r}_3|n,\beta ;t)\right\} \\
   \times F_1(x_1;t)F_1(x_2;t)F_1(x_3;t).
 \end{multline}\\[-8mm]
 Next,
 we introduce the operator that can be obtained by varying of the mean field collision operator near the nonequilibrium distribution $F_1(x_1;t)$:\vspace{-2mm}
 \begin{equation*}%\label{e4.10}
  \delta\left(\int d x_3 \,iL(1,3)\,F_1(x_3;t)\,F_1(x_1;t)\right)
  =\int dx_3 \,iL(1,3)\,F_1(x_3;t)\,\delta F_1(x_1;t)
  =L(x_1;t)\,\delta F_1(x_1;t).
 \end{equation*}\\[-5mm]
 Then the equation (\ref{e4.9}) using the operator $L(x_1;t)$ is represented in in the form of\vspace{-3mm}
 \begin{multline*}%\label{e4.11}
  \left(\frac{\partial }{\partial t}+iL_2+L(x_1,x_2;t)+\varepsilon\right) F_2(x_1,x_2;t)\\
  =
  -\left(\frac{\partial }{\partial t}+iL_2+L(x_1,x_2;t)\right)
  g_2(\vec{r}_1,\vec{r}_2|n,\beta ;t)F_1(x_1;t)F_1(x_2;t),
 \end{multline*}\\[-5mm]
 where we find the formal solution for the fundamental two-particle function distribution:\vspace{-3mm}
 \begin{multline} \label{e4.12}
  F_2(x_1,x_2;t)=-\int_{-\infty}^{t} dt' e^{\varepsilon(t'-t)} U(t,t')\\[-1mm]
  \times\left(\frac{\partial }{\partial t'}+iL_2+L(x_1,x_2;t')\right) g_2(\vec{r}_1,\vec{r}_2|n,\beta ;t')
    F_1(x_1;t')F_1(x_2;t'),
 \end{multline}\\[-5mm]
 where $U(t,t')$ is an evolution operator:\vspace{-4mm}
 \begin{align*}%\label{e4.13}
  U(t,t')&=\exp_{+}\left(-\int_{t'}^{t} dt''\left(iL_2+L(x_1,x_2;t'')\right)\right),\\
  &L(x_1,x_2;t)=L(x_1;t)+L(x_2;t).\nonumber
 \end{align*}\\[-7mm]
 As a result,
 we obtain the expression for the irreducible quasi-equilibrium two-particle distribution functions $F_2(x_1,x_2;t)$ in generalized polarization approximation.
 Substitute this expression~(\ref{e4.12}) into the first equation of the chain~(\ref{e4.4}),
 then we get:\vspace{-4mm}
 \begin{multline} \label{e4.14}
  \left(\frac{\partial }{\partial t}+iL(1)\right) F_1(x_1;t)
  +\int dx_2 \,iL(1,2)\,g_2(\vec{r}_1,\vec{r}_2|n,\beta;t)\,F_1(x_1;t)\,F_1(x_2;t)
    \\
  =\int dx_2 \int_{-\infty}^{t}dt' e^{\varepsilon(t'-t)} iL(1,2)U(t,t') \\
  \times\left(\frac{\partial }{\partial t'}+iL_2+L(x_1,x_2;t')\right)
  g_2(\vec{r}_1,\vec{r}_2|n,\beta;t')    F_1(x_1;t')F_1(x_2;t')
 \end{multline}\\[-5mm]
 is a generalized kinetic equation for a nonequilibrium one-particle function distribution with the non-Markov collision integral in the generalized polarization approximation.
 It should be noted that the presence in the integral collisions~(\ref{e4.14}) of the mean field collision operator $L(x_1,x_2;t)$ indicates that it takes into account the collective effects.
 In general,
 the analysis of the integral of collisions in~(\ref{e4.14}) is a rather complex mathematical task.
 Obviously,
 for each physical model of a system of particles,
 or of the nonequilibrium state,
 the integral of collisions in~(\ref{e4.14}),
 or the expression for $F_2(x_1,x_2;t)$~(\ref{e4.12}) can be significantly simplified.
 In addition,
 for further application of the obtained kinetic equations,
 in particular,
 their solution by the Chapman--Enskog method and construction of the corresponding equations of hydrodynamics,
 it is necessary to set the algorithm for calculating the pair coordinate quasi-equilibrium distribution function $g_{2}(\vec{r},\vec{r}'|n,\beta ;t)$.

 In the next section,
 the method of collective variables Yukhnovskii~\cite{19c,18c,20c,hlush,YukhHlush} will be used to calculate the pair quasi-equilibrium distribution function and higher-order distribution functions.

 \vspace{-1mm}
 \section{Calculation of the statistical sum of the quasi-equilibrium distribution and coordinate functions of the distribution $g_{s}((\vec{r})^{s}|n,\beta;t)$ by the method of collective variables}
 \vspace{-1mm}

 For calculations of the even coordinate quasi-equilibrium distribution function $g_{2}(\vec{r},\vec{r}'|n,\beta ;t)$ and higher correlation functions $g_{s}((\vec{r})^{s}|n,\beta ;t)$ use the statistical sum of the quasi-equilibrium distribution:\vspace{-2mm}
 \begin{equation*} %\label{e.401}
  Z_{rel}(t)=\int d\Gamma_{N}(x)\,
  \exp\left\{-\int d\vec{r}\,\beta(\vec{r},t)\,\hat{\varepsilon}_{int}(\vec{r})-\int dx \,a(x,t)\,\hat{n}(x)\right\} .
 \end{equation*}\\[-5mm]
 Correlation functions $g_{s}((\vec{r})^{s}|n,\beta ;t)$ are determined by variational derivatives on the parameter $a(x,t)$ of the corresponding order from the statistical sum $Z_{rel}(t)$:\vspace{-2mm}
 \begin{equation*}    %   \label{e.402}
  g_{s}((\vec{r})^{s}|n,\beta ;t)=\frac{1}{n(\vec{r}_{1},t)\ldots n(\vec{r}_{s},t)}
  \int d\vec{p}_{1}\ldots\int d\vec{p}_{s}\,
  \frac{\delta}{\delta a(\vec{p}_{1},\vec{r}_{1},t)}\ldots\frac{\delta}{\delta a(\vec{p}_{s},\vec{r}_{s},t)}\ln Z_{rel}(t).
 \end{equation*}\\[-5mm]
 The calculation of the statistical sum of the quasi-equilibrium distribution will be performed using the method of collective variables~\cite{19c,18c,20c,hlush,YukhHlush}.
 In doing so,
 we will take into account the nature of short-range and long-range interactions of particles,
 therefore $Z_{rel}(t)$ is given in the form:\vspace{-2mm}
 \begin{equation*}   %    \label{e.403}
  Z_{rel}(t)=\int d\Gamma_{N}(x)\,
  \exp\left\{-\int d\vec{r}\,\beta(\vec{r},t)\,\hat{\varepsilon}_{int}^{sr}(\vec{r})
  -\int d\vec{r}\,\beta(\vec{r},t)\,\hat{\varepsilon}_{int}^{lr}(\vec{r})
  -\int dx \,a(x,t)\,\hat{n}(x)\right\} .
 \end{equation*}\\[-10mm]

 Next,
 assuming that for the far-reaching part of the potential $\Phi_{attr}^{lr}(|{\bf r}_{lj}|)$ there is a Fourier image $\nu(\vec{k})$ in the space of wave vectors $\vec{k}$,
 the far-reaching part of $\int d\vec{r}\,\beta(\vec{r},t)\,\hat{\varepsilon}_{int}^{lr}(\vec{r})$ will be presented in the form of
 \begin{equation} \label{e.404}
  \int d\vec{r}\,\beta(\vec{r},t)\,\hat{\varepsilon}_{int}^{lr}(\vec{r})
  =\frac{1}{2V^{2}}\sum_{\vec{q}}\sum_{\vec{k}}\beta_{-\vec{q}}(t)\,\nu(\vec{k})
  \left(\hat{\rho}_{\vec{q}+\vec{k}}\,\hat{\rho}_{-\vec{k}}-\hat{\rho}_{\vec{q}}\right),
 \end{equation}\\[-5mm]
 where
 \begin{equation*} %   \label{e.405}
  \hat{\rho}_{\vec{k}}=\sum_{j=1}^{N}\mbox{e}^{-i{\vec{k}} \vec{r}_{j}}
 \end{equation*}\\[-3mm]
 is the Fourier component of the particle number density.
 Taking into account~(\ref{e.404}),
 the statistical sum of the quasi-equilibrium distribution is given in the form:\vspace{-2mm}
 \begin{align}       \label{e.406}
  Z_{rel}(t)&=\int d\Gamma_{N}(x)
  \exp\left\{-\int d\vec{r}\,\beta(\vec{r},t)\,\hat{\varepsilon}_{int}^{sr}(\vec{r})
             -\int dx\,a(x,t)\,\hat{n}(x)\right\}\nonumber\\
  &\quad\times\exp\left\{-\frac{1}{2V^{2}}\sum_{\vec{q}}
  \sum_{\vec{k}}\beta_{-\vec{q}}(t)\,\nu(\vec{k})
  \left(\hat{\rho}_{\vec{q}+\vec{k}}\hat{\rho}_{-\vec{k}}-\hat{\rho}_{\vec{q}}\right)\right\}.
 \end{align}\\[-4mm]
 Next,
 enter the Jacobian transition\vspace{-2mm}
 \begin{equation*}  %  \label{e.407}
  \hat{J}(\rho)=\delta (\hat{\rho}-\rho)=
  \prod_{\vec{k}}\delta (\hat{\rho}_{\vec{k}}-\rho_{\vec{k}})
 \end{equation*}\\[-5mm]
 to the collective variables $\rho_{\vec{k}}$,
 the statistical sum~(\ref{e.406}) is written as:\vspace{-2mm}
 \begin{equation*}    %   \label{e.408}
  Z_{rel}(t)=Z_{rel}^{sh}(t)\int d\rho\, \langle \hat{J}(\rho)\rangle^{t}_{rel,sh}
  \exp\left\{-\frac{1}{2V^{2}}\sum_{\vec{q}}\sum_{\vec{k}}\beta_{-\vec{q}}(t)\,
  \nu(\vec{k})\left(\rho_{\vec{q}+\vec{k}}\rho_{-\vec{k}}-\rho_{\vec{q}}\right)\right\},
 \end{equation*}\\[-5mm]
 where $d\rho=\prod_{\vec{k}}d\rho_{\vec{k}}$ and\vspace{-3mm}
 \begin{equation*}     %  \label{e.409}
  \langle \hat{J}(\rho)\rangle^{t}_{rel,sh}=\int d\Gamma_{N}(x) \delta (\hat{\rho}-\rho)\,\rho_{rel,sh}(x^{N};t)
 \end{equation*}\\[-5mm]
 is the average Jacobian transition to collective variables by quasi-equilibrium distribution,
 which takes into account only short-range repulsive and attractive interactions between particles\vspace{-2mm}
 \begin{equation}       \label{e.410}
  \rho_{rel,sh}(x^{N};t)=\frac{1}{Z_{rel}^{sh}(t)}
  \exp\left\{-\int d\vec{r}\,\beta(\vec{r},t)\,\hat{\varepsilon}_{int}^{sr}(\vec{r})
  -\int dx \,a(x,t)\,\hat{n}(x)\right\},
 \end{equation}\\[-4mm]
 the statistical sum of which has the following structure:\vspace{-2mm}
 \begin{equation*}   %    \label{e.411}
  Z_{rel}^{sh}(t)
  =\int d\Gamma_{N}(x)
  \exp\left\{-\int d\vec{r}\,\beta(\vec{r},t)\,\hat{\varepsilon}_{int}^{sr}(\vec{r})
  -\int dx\, a(x,t)\,\hat{n}(x)\right\}.
 \end{equation*}\\[-9mm]

 Next,
 we apply the integral representation for the $\delta$-function,
 then $\hat{J}(\rho)$ is given as\vspace{-2mm}
 \begin{equation*}   % \label{eq3.1}
  \hat{J}(\rho)=\int d\omega \exp\Big\{ -i\pi \sum_{\vec{k}}\omega_{\vec{k}}(\hat{\rho}_{\vec{k}}-\rho_{\vec{k}})\Big\}.
 \end{equation*}\\[-5mm]
 Using the cumulative schedule~\cite{20c,hlush,YukhHlush} for $\langle \hat{J}(\rho)\rangle^{t}_{rel,sh}=W(\rho;t)$,
 we get:\vspace{-2mm}
 \begin{align}    \label{eq3.2}
  W(\rho;t)&=\langle \hat{J}(\rho)\rangle^{t}_{rel,sh}=\int d\Gamma_{N}\rho_{rel,sh} (x^{N};t)\hat{J}(\rho)\nonumber\\
  &=\int d\omega \exp \bigg\{-i\pi \sum_{\vec{k}}\omega_{\vec{k}}\bar{\rho}_{\vec{k}}-\frac{\pi^{2}}{2} \sum_{{\vec{k}}_{1},{\vec{k}}_{2}}\mathfrak{M}_{2}({\vec{k}}_{1},{\vec{k}}_{2};t)\omega_{{\vec{k}}_{1}}
  \omega_{{\vec{k}}_{2}}\bigg\}  \exp\bigg\{ \sum_{n\geq 3}D_{n}(\omega ;t)\bigg\},
 \end{align}\\[-5mm]
 where\vspace{-3mm}
 \begin{align*}
  \bar{\rho}_{\vec{k}}&=\rho_{\vec{k}}-\langle\hat{\rho}_{\vec{k}}\rangle_{rel,sh}^{t}, &
    &d\omega=\Pi_{\vec{k}}d\omega_{\vec{k}}^{r}d\omega_{\vec{k}}^{s},            \\
  \omega_{\vec{k}}&=\omega_{\vec{k}}^{r}-i\omega_{\vec{k}}^{s},    & &
  \omega_{-\vec{k}}=\omega_{\vec{k}}^{*},
 \end{align*}
 \begin{equation*}   % \label{Eq3.3}
  D_{n}(\omega ;t)=\frac{(-i\pi)^{n}}{n!}
  \sum_{{\vec{k}}_{1},\ldots,{\vec{k}}_{n}}
  \mathfrak{M}_{n}({\vec{k}}_{1},\ldots,{\vec{k}}_{n};t)\,\omega_{{\vec{k}}_{1}}\ldots  \omega_{{\vec{k}}_{n}}.
 \end{equation*}\\[-3mm]
 Quasi-equilibrium cumulative averages of $n$-order\vspace{-2mm}
 \begin{equation*}  %  \label{Eq3.4}
  \mathfrak{M}_{n}({\vec{k}}_{1},\ldots,{\vec{k}}_{n};t)=
  \langle \hat{\rho}_{{\vec{k}}_{1}}\ldots\hat{\rho}_{{\vec{k}}_{n}}\rangle_{rel,sh}^{t,c}
 \end{equation*}\\[-7mm]
 calculated with the relevant distribution function with short-range interparticle interaction~(\ref{e.410}),
 where the superscript $c$ means the cumulative average.
 It is important to note that in~(\ref{e.002}) we have separated the contributions from the short-term ones and long-range interactions.
 Short-term interactions are taken into account in the relevant distribution~(\ref{e.410})
 (which can be considered as basic),
 and long-range interactions are presented through collective variables.

 For further calculation,
 the structural function $W(\rho;t)$ is given in the form:\vspace{-2mm}
 \begin{align}    \label{eq3.6}
  W(\rho;t)&=\int d\omega \exp\bigg\{ -i\pi \sum_{\vec{k}}
 \omega_{\vec{k}}\bar{\rho}_{\vec{k}} -\frac{\pi^{2}}{2}\sum_{{\vec{k}}_{1},{\vec{k}}_{2}}
    \mathfrak{M}_{2}({\vec{k}}_{1},{\vec{k}}_{2};t)\omega_{{\vec{k}}_{1}}
    \omega_{{\vec{k}}_{2}}\bigg\}          \nonumber\\
  &\quad \times\left(1+B+\frac{1}{2!}B^{2}+\frac{1}{3!}B^{3}+\ldots+\frac{1}{n!}B^{n}+\dots\right),
 \end{align}\\[-5mm]
 where $B=\sum_{n\geq 3}D_{n}(\omega ;t)$.
 If in the schedule in a number of exponents (\ref{eq3.6}),
 ie $\exp\{\sum_{n\geq 3}D_{n}(\omega ;t)\}$,
 save only the first term,
 equal to one,
 we obtain the Gaussian approximation for $W(\rho;t)$:\vspace{-2mm}
 \begin{equation}    \label{eq3.7}
  W^{G}(\rho;t)=\int d\omega
  \exp \bigg\{ i\pi \sum_{\vec{k}}\omega_{\vec{k}}\bar{\rho}_{\vec{k}}
  -  \frac{\pi^{2}}{2} \sum_{{\vec{k}}_{1},{\vec{k}}_{2}}
  \mathfrak{M}_{2}({\vec{k}}_{1},{\vec{k}}_{2};t)\,
  \omega_{{\vec{k}}_{1}} \omega_{{\vec{k}}_{2}}\bigg\},
 \end{equation}\\[-5mm]
 where the quasi-equilibrium cumulative average density--density has the form:\vspace{-2mm}
 \begin{equation*}
  \mathfrak{M}_{2}({\vec{k}}_{1},{\vec{k}}_{2};t)
  =
  \langle \hat{\rho}_{{\vec{k}}_{1}}\hat{\rho}_{{\vec{k}}_{2}}\rangle_{rel,sh}^{t,c}
  =
  \langle \hat{\rho}_{\vec{k}} \hat{\rho}_{-\vec{k}}\rangle_{rel,sh}^{t}
  -
  \langle \hat{\rho}_{\vec{k}} \rangle_{rel,sh}^{t}\langle  \hat{\rho}_{-\vec{k}}\rangle_{rel,sh}^{t}.
 \end{equation*}

 To integrate for $d\omega$ in (\ref{eq3.7}) it is necessary to give expression in the exponent to the quadratic diagonal form for $\omega_{\vec{k}}$.
 In this regard,
 you need to find your values by solving the equation:\vspace{-2mm}
 \begin{equation*}%\label{Eq3.10}
  \det\big|\tilde{\mathfrak{M}}_{2}({\vec{k}}_{1},{\vec{k}}_{2};t)-\tilde{E}({\vec{k}}_{1},{\vec{k}}_{2};t)\big|=0,
 \end{equation*}\\[-7mm]
 $\tilde{E}({\vec{k}}_{1},{\vec{k}}_{2};t)$  is a diagonal matrix.
 With this in mind,
 we obtain:\vspace{-2mm}
 \begin{equation}\label{eq3.11}
  W^{G}(\rho;t) = \int d\omega\,
  \exp \bigg\{-i\pi \sum_{\vec{k}} \bar{\rho}_{\vec{k}}
 \omega_{\vec{k}}
- \frac{\pi^{2}}{2}\sum_{{\vec{k}}}
 E({\vec{k}};t)\,\omega_{\vec{k}} \omega_{-\vec{k}}\bigg\}.
 \end{equation}\\[-5mm]
 The subintegral expression in (\ref{eq3.11}) is a quadratic function $\tilde{\omega}_{\vec{k}}$,
 therefore,
 performing integration for $d\omega_{\vec{k}}$,
 for a structural function in the Gaussian approximation $W^{G}(a;t)$ we obtain:\vspace{-2mm}
 \begin{equation*}%\label{Eq3.12}
  W^{G}(\rho;t)
  =
  Z(t)  \exp \left\{ -\frac{1}{2}\sum_{\vec{k}} E^{-1}({\vec{k}};t)\,
  \bar{\rho}_{{\vec{k}}}\,\bar{\rho}_{-{\vec{k}}}\right\},
 \end{equation*}\\[-5mm]
 where
 \[
  Z(t)= \exp \left\{-\frac{1}{2} \sum_{\vec{k}} \ln\pi \det \tilde{E}({\vec{k}};t)\right\}.
 \]\\[-2mm]
 The structural function $W^{G}(\rho;t)$ in the Gaussian approximation makes it possible to calculate full structural function~(\ref{eq3.2}) in higher approximations by Gaussian moments~\cite{hlush,YukhHlush}:\vspace{-2mm}
 \begin{equation*}  %\label{eq3.14}
   W(\rho;t)=\bar{W}^{G}(\rho;t)\exp \left\{\sum_{n\geq 3}\langle \tilde{D}_{n}(\rho ;t)\rangle_{G}\right\},
 \end{equation*}\\[-5mm]
 where
 \begin{equation*} %\label{eq3.014}
  \bar{W}^{G}(\rho;t)=Z(t)\exp \left\{ -\frac{1}{2}\sum_{\vec{k}} E^{-1}({\vec{k}};t)\,
  \bar{\rho}_{{\vec{k}}}\,\bar{\rho}_{-{\vec{k}}}\right\}
 \end{equation*}\\[-3mm]
 and $\langle \tilde{D}_{n}(\rho ;t)\rangle_{G}$ approximate as follows:\vspace{-2mm}
 \begin{align*}
  \langle\tilde{D}_{3}(\rho ;t)\rangle_{G} &= \langle \bar{D}_{3}(\rho ;t)\rangle_{G},  \\
  \langle\tilde{D}_{4}(\rho ;t)\rangle_{G} &= \langle \bar{D}_{4}(\rho ;t)\rangle_{G},  \\
  \langle\tilde{D}_{6}(\rho ;t)\rangle_{G} &= \langle \bar{D}_{6}(\rho ;t)\rangle_{G}-
   \frac{1}{2}\langle\bar{D}_{3}(\rho ;t)\rangle_{G}^{2},   \\
  \langle\tilde{D}_{8}(\rho ;t)\rangle_{G} &= \langle \bar{D}_{8}(\rho ;t)\rangle_{G}-
  \langle\bar{D}_{3}(\rho;t)\rangle_{G}\langle \bar{D}_{5}(\rho ;t)\rangle_{G} -
   \frac{1}{2}\langle \bar{D}_{4}(\rho ;t)\rangle_{G}^{2},   \\
  \langle\tilde{D}_{n}(\rho;t)\rangle_{G} &=\frac{1}{\bar{W}^{G}(\rho;t)}\sum_{{\vec{k}}_{1},\ldots,{\vec{k}}_{n}}
   \bar{\mathfrak{M}}_{n}({\vec{k}}_{1},\ldots,{\vec{k}}_{n};t)
   \frac{1}{(i\pi)^{n}}\frac{\delta^{n}}{\delta \bar{\rho}_{{\vec{k}}_{1}}\ldots
   \delta \bar{\rho}_{{\vec{k}}_{n}}}\bar{W}^{G}(\rho;t).
 \end{align*}\\[-4mm]
 $\langle\tilde{D}_{n}(\rho;t)\rangle_{G}$ are renormalized $n$-quasi-equilibrium cumulative averages for variables $\bar{\rho}_{\vec{k}}$ higher orders.\vspace{-2mm}
 \begin{equation*}%\label{e.4088}
  Z_{rel}(t)=Z_{rel}^{sh}(t)\int d\rho \,
  W_{\beta}(\rho;t)\,
  \bar{W}^{G}(\rho;t)\exp \bigg\{\sum_{n\geq 3}\langle \tilde{D}_{n}(\rho ;t)\rangle_{G}\bigg\},
 \end{equation*}\\[-5mm]
 where
 \begin{equation*} %\label{eq3.06}
   W_{\beta}(\rho;t)=\exp\left(- \frac{1}{2V^2}\sum_{\vec{q}}\sum_{\vec{k}}
   \beta_{-{\vec{q}}}(t)\nu({\vec{k}})    \big(\rho_{{\vec{q}}+{\vec{k}}}  \rho_{-{\vec{k}}} - \rho_{{\vec{q}}}\big)\right).
 \end{equation*}
 In the Gaussian approximation of the collective variables for $W_{\beta}(\rho;t)$ quasi-equilibrium distribution function will look like this:\vspace{-2mm}
 \begin{equation*}  %\label{e.4089}
  Z_{rel}^{G}(t)=Z_{rel}^{sh}(t)\int (d\rho)  \,W_{\beta}(\rho;t)\,\bar{W}^{G}(\rho;t).
 \end{equation*}\\[-5mm]
 In this approximation for the pair quasi-equilibrium function of particles,
 we obtain:\vspace{-2mm}
 \begin{equation}       \label{e.4020}
  g_{2}(\vec{r}_{1},\vec{r}_{2}|n,\beta ;t)=\frac{1}{n(\vec{r}_{1},t)\,n(\vec{r}_{2},t)}
  \int d\vec{p}_{1}\int d\vec{p}_{2}\,
  \frac{\delta}{\delta a(\vec{p}_{1},\vec{r}_{1},t)}
  \frac{\delta}{\delta a(\vec{p}_{},\vec{r}_{2},t)}
  \ln\bigg(Z_{rel}^{sh}(t)
 \end{equation}
 \[
  \times\exp\bigg\{-\frac{1}{2}\sum_{\vec{k}}\ln\pi\det\tilde{E}({\vec{k}};t)\bigg\}V(\beta,n;t)\bigg),
 \]\\[-5mm]
 where\vspace{-2mm}
 \[
  V(\beta,n;t)=\int (d\rho)
  \exp\bigg\{-\frac{1}{2}\sum_{\vec{k}} E^{-1}({\vec{k}};t)\,\bar{\rho}_{{\vec{k}}}\,\bar{\rho}_{-{\vec{k}}}\bigg\}
  \exp\bigg(\!-\frac{1}{2V^2}\sum_{\vec{q}}\sum_{\vec{k}}
   \beta_{-{\vec{q}}}(t)\,\nu({\vec{k}})   \big(\rho_{{\vec{q}}+{\vec{k}}}  \rho_{-{\vec{k}}} - \rho_{{\vec{q}}}\big)\bigg).
 \]\\[-5mm]
 To integrate the collective variables in $V(\beta,n;t)$ in the first exponent under the integral we write $\bar{\rho}_{{\vec{k}}}=\rho_{{\vec{k}}}-\langle\hat{\rho}_{{\vec{k}}}\rangle_{rel,sh}^{t}$.
 Then we get\vspace{-3mm}
 \begin{multline*}
  -\frac{1}{2}\sum_{\vec{k}} E^{-1}({\vec{k}};t)
  \bar{\rho}_{{\vec{k}}}\bar{\rho}_{-{\vec{k}}}=-\frac{1}{2}\sum_{\vec{k}} E^{-1}({\vec{k}};t)
  \rho_{{\vec{k}}}\rho_{-{\vec{k}}}\\[-1mm]
  +\sum_{\vec{k}} E^{-1}({\vec{k}};t)
  \rho_{{\vec{k}}}\langle\hat{\rho}_{-{\vec{k}}}\rangle_{rel,sh}^{t}
  -\frac{1}{2}\sum_{\vec{k}} E^{-1}({\vec{k}};t)\langle\hat{\rho}_{{\vec{k}}}\rangle_{rel,sh}^{t}\langle\hat{\rho}_{-{\vec{k}}}\rangle_{rel,sh}^{t}.
 \end{multline*}\\[-5mm]
 Then in the second exponent under the integral,
 you need the first term to the quadratic form:\vspace{-2mm}
 \[
  -\frac{1}{2V^2}\sum_{\vec{q}}\sum_{\vec{k}}\beta_{-{\vec{q}}}(t)\,\nu({\vec{k}})   \big(\rho_{{\vec{q}}+{\vec{k}}}  \rho_{-{\vec{k}}} - \rho_{{\vec{q}}}\big)
  =
  -\frac{1}{2V^2}\sum_{\vec{k}}\lambda(\vec{k},t)\,\rho_{{\vec{k}}}\,\rho_{-{\vec{k}}}
  +\frac{1}{2V^2}\sum_{\vec{k}'}\nu({\vec{k}'})\sum_{\vec{k}}\beta_{-{\vec{k}}}(t)\,\rho_{{\vec{k}}},
 \]\\[-4mm]
 where $\lambda({\vec{k}}_{1},{\vec{k}}_{2};t)$ there are eigenvalues from the equation:\vspace{-2mm}
 \begin{equation*}  %\label{Eq3.1001}
   \det\big|\tilde{\mathfrak{B}}({\vec{k}}_{1},{\vec{k}}_{2};t)-\tilde{\lambda}({\vec{k}}_{1},{\vec{k}}_{2};t)\big|=0,
 \end{equation*}\\[-7mm]
 in which $\mathfrak{B}({\vec{k}}_{1},{\vec{k}}_{2};t)=\beta_{{\vec{k}}_{1}-{\vec{k}}_{2}}(t)\nu({\vec{k}}_{1})$.

 Given the transformation data in the exponents of $V(\beta,n;t)$,
 you can submit it as\vspace{-3mm}
 \begin{multline*}
   V(\beta,n;t)=\int (d\rho) \exp \bigg\{ -\frac{1}{2}\sum_{\vec{k}} G({\vec{k}};t)\Big(\rho_{{\vec{k}}}-\frac{1}{2}R({\vec{k}};t)\Big)^{2}\bigg\} \\[-1mm]
   \times\exp \bigg\{ -\frac{1}{2}\sum_{\vec{k}}\Big(E^{-1}({\vec{k}};t)\langle\hat{\rho}_{{\vec{k}}}\rangle_{rel,sh}^{t}\langle\hat{\rho}_{-{\vec{k}}}\rangle_{rel,sh}^{t}-
   \frac{1}{4}G({\vec{k}};t)R^{2}({\vec{k}};t)\Big)\bigg\}
 \end{multline*}\\[-5mm]
 where\vspace{-2mm}
 \begin{align*}
  G({\vec{k}};t)&=E^{-1}({\vec{k}};t)+\frac{1}{2V^2}\lambda (\vec{k},t),\\
  R({\vec{k}};t)&=\frac{1}{1+\frac{1}{2V^2}\lambda (\vec{k},t)E({\vec{k}};t)}\bigg(\langle\hat{\rho}_{{\vec{k}}}\rangle_{rel,sh}^{t}
  +\frac{1}{2V^2}\sum_{\vec{k}'}\nu({\vec{k}'})\beta_{{\vec{k}}}(t)\bigg).
 \end{align*}\\[-5mm]
 After integration by collective variables,
 we finally get\vspace{-2mm}
 \begin{multline*}
  V(\beta,n;t)=\exp\bigg\{-\frac{1}{2} \sum_{\vec{k}} \ln\,\det \tilde{G}({\vec{k}};t)\bigg\}\\[-1mm]
  \times\exp\bigg\{ -\frac{1}{2}\sum_{\vec{k}}\bigg( E^{-1}({\vec{k}};t)\langle\hat{\rho}_{{\vec{k}}}\rangle_{rel,sh}^{t}\langle\hat{\rho}_{-{\vec{k}}}\rangle_{rel,sh}^{t}-
  \frac{1}{4}G({\vec{k}};t)R^{2}({\vec{k}};t)\bigg)\bigg\}.
 \end{multline*}\\[-5mm]
 After substituting $V(\beta,n;t)$ in (\ref{e.4020}),
 we get:\vspace{-3mm}
 \begin{align*}       %\label{e.4022}
  g_{2}(\vec{r}_{1},\vec{r}_{2}|n,\beta ;t)&=g_{2}^{sh}(\vec{r}_{1}\,\vec{r}_{2}|n,\beta ;t)
  -\frac{1}{n(\vec{r}_{1},t),n(\vec{r}_{2},t)} \nonumber\\
   &\quad\times \int\!\! d\vec{p}_{1}\int \!\!d\vec{p}_{2}\,
   \frac{\delta}{\delta a(\vec{p}_{1},\vec{r}_{1},t)}\,
   \frac{\delta}{\delta a(\vec{p}_{2},\vec{r}_{2},t)}
   \bigg(\frac{1}{2}\sum_{\vec{k}}\ln\pi\det\tilde{E}({\vec{k}};t)
   +\frac{1}{2}\sum_{\vec{k}}\ln\det\tilde{G}({\vec{k}};t) \nonumber\\
   &\qquad\qquad+\frac{1}{2}\sum_{\vec{k}}
  \bigg(E^{-1}({\vec{k}};t)\langle\hat{\rho}_{{\vec{k}}}\rangle_{rel,sh}^{t}\langle\hat{\rho}_{-{\vec{k}}}\rangle_{rel,sh}^{t}-
   \frac{1}{4}G({\vec{k}};t)R^{2}({\vec{k}};t)\bigg)
  \bigg),
 \end{align*}\\[-5mm]
 where $g_{2}^{sh}(\vec{r}_{1},\vec{r}_{2}|n,\beta ;t)$ is an even quasi-equilibrium coordinate function of particle distribution with short-range interaction.
 The following terms are complex in structure and are related to the functional derivatives
 $\frac{\delta}{\delta a(\vec{p}_{1},\vec{r}_{1},t)}\frac{\delta}{\delta a(\vec{p}_{},\vec{r}_{2},t)}$,
 acting on $\langle\hat{\rho}_{{\vec{k}}}\rangle_{rel,sh}^{t}$ and $\mathfrak{M}_{2}({\vec{k}}_{1},{\vec{k}}_{2};t)$,
 through which the functions $E({\vec{k}};t)$, $G({\vec{k}};t)$ , $R({\vec{k}};t)$.
 The results are as follows:\vspace{-3mm}
 \begin{align*}
  \frac{\delta}{\delta a(\vec{p}_{2},\vec{r}_{2},t)}&
  \langle\hat{\rho}_{{\vec{k}}}\rangle_{rel,sh}^{t}
  =\langle\hat{\rho}_{{\vec{k}}}\rangle_{rel,sh}^{t}
   \langle\hat{n}(x_{2})\rangle_{rel,sh}^{t}-\langle\hat{\rho}_{{\vec{k}}}\hat{n}(x_{2})\rangle_{rel,sh}^{t},\\
  \frac{\delta}{\delta a(\vec{p}_{1},\vec{r}_{1},t)}&\frac{\delta}{\delta a(\vec{p}_{2},\vec{r}_{2},t)}\langle\hat{\rho}_{{\vec{k}}}\rangle_{rel,sh}^{t}
=2\langle\hat{\rho}_{{\vec{k}}}\rangle_{rel,sh}^{t} \langle\hat{n}(x_{1})\rangle_{rel,sh}^{t}
\langle\hat{n}(x_{2})\rangle_{rel,sh}^{t}-\langle\hat{n}(x_{1})\rangle_{rel,sh}^{t}\langle\hat{\rho}_{{\vec{k}}}\hat{n}(x_{2})\rangle_{rel,sh}^{t}
\\
&-\langle\hat{\rho}_{{\vec{k}}}\rangle_{rel,sh}^{t} \langle\hat{n}(x_{1})\hat{n}(x_{2})\rangle_{rel,sh}^{t}-\langle\hat{n}(x_{2})\rangle_{rel,sh}^{t}\langle\hat{\rho}_{{\vec{k}}}\hat{n}(x_{1})\rangle_{rel,sh}^{t}
+\langle\hat{\rho}_{{\vec{k}}}\hat{n}(x_{1})\hat{n}(x_{2})\rangle_{rel,sh}^{t},\\
\frac{\delta}{\delta a(\vec{p}_{2},\vec{r}_{2},t)}&
\langle\hat{\rho}_{{\vec{k}}}\hat{\rho}_{-{\vec{k}}}\rangle_{rel,sh}^{t}=\langle\hat{\rho}_{{\vec{k}}}\hat{\rho}_{-{\vec{k}}}\rangle_{rel,sh}^{t}
\langle\hat{n}(x_{2})\rangle_{rel,sh}^{t}
-\langle\hat{\rho}_{{\vec{k}}}\hat{\rho}_{-{\vec{k}}}\hat{n}(x_{2})\rangle_{rel,sh}^{t},\\
\frac{\delta}{\delta a(\vec{p}_{1},\vec{r}_{1},t)}&\frac{\delta}{\delta a(\vec{p}_{2},\vec{r}_{2},t)}\langle\hat{\rho}_{{\vec{k}}}\hat{\rho}_{-{\vec{k}}}\rangle_{rel,sh}^{t}
=
2\langle\hat{\rho}_{{\vec{k}}}\hat{\rho}_{-{\vec{k}}}\rangle_{rel,sh}^{t}
\langle\hat{n}(x_{1})\rangle_{rel,sh}^{t}\langle\hat{n}(x_{2})\rangle_{rel,sh}^{t}\\
&-\langle\hat{\rho}_{{\vec{k}}}\hat{\rho}_{-{\vec{k}}}\rangle_{rel,sh}^{t}\langle\hat{n}(x_{1})\hat{n}(x_{2})\rangle_{rel,sh}^{t}
-\langle\hat{n}(x_{1})\rangle_{rel,sh}^{t}\langle\hat{\rho}_{{\vec{k}}}\hat{\rho}_{-{\vec{k}}}\hat{n}(x_{2})\rangle_{rel,sh}^{t}\\
&-\langle\hat{n}(x_{2})\rangle_{rel,sh}^{t}\langle\hat{\rho}_{{\vec{k}}}\hat{\rho}_{-{\vec{k}}}\hat{n}(x_{1})\rangle_{rel,sh}^{t}
+\langle\hat{\rho}_{{\vec{k}}}\hat{\rho}_{-{\vec{k}}}\hat{n}(x_{1})\hat{n}(x_{2})\rangle_{rel,sh}^{t}.
\end{align*}\\[-11mm]

 After integrating by impulses $\int d\vec{p}_{1}\int d\vec{p}_{2}$ we get the expression for $g_{2}(\vec{r}_{1},\vec{r}_{2}|n,\beta ;t)$ through one-,
 two-,
 three-,
 four-particle coordinate quasi-equilibrium distribution functions for particles with short-range interaction.
 In addition,
 $g_{2}(\vec{r}_{1},\vec{r}_{2}|n,\beta ;t)$ also depends on the long-range part of the interaction $\nu({\vec{k}})$ due to certain renormalizations $G({\vec{k}};t)$,
 and is also a function of the local inverse temperature $\beta({\vec{r}};t)$.
 Thus,
 we obtain the expression for the quasi-equilibrium pair coordinate function of the particle distribution in the Gaussian approximation $W^{G}(\rho;t)$ (chaotic phase type approximation).
 Higher approximations will be related to quasi-equilibrium renormalized cumulatives
 $\langle \tilde {D}_{n}(\rho;t)\rangle_{G}$,
 which will also be expressed in terms of one-,
 two-,
 three-,
 four- and higher-order particle coordinate quasi-equilibrium distribution functions for particles with short-range interaction.
 And so we come to the problem of calculating one-,
 two-,
 three-,
 four- and higher-order particle coordinate quasi-equilibrium distribution functions for particles with short-range interaction.

\vspace{-2.5mm}
 \section{Conclusions}\vspace{-2mm}

 Within the concept of the coordinated description of kinetic and hydrodynamic processes,
 generalized kinetic equations are obtained for the nonequilibrium one-particle distribution function in the approximation of ``paired'' collisions and in the polarization approximation when the interaction through the third particle is taken into account.
 In this case,
 the contributions from the short-range and long-range nature of the interaction of particles are distinguished in the collision integrals.
 The resulting collision integrals include a quasi-equilibrium coordinate distribution function $g_{2}(\vec{r}_{1},\vec{r}_{2}|n,\beta ;t)$ as a function of nonequilibrium densities of the number of particles $n({\vec{r}};t)$ and inverse temperature $\beta({\vec{r}};t)$ and describes multiparticle correlations.
 $g_{2}(\vec{r}_{1},\vec{r}_{2}|n,\beta ;t)$ is of independent interest,
 as it may be related to the corresponding dynamic structural factor quasi-balanced state of the system.
 The calculation of $g_{2}(\vec{r}_{1},\vec{r}_{2}|n,\beta ;t)$ is one of the important problems~\cite{Karkheck3,Polewczak1,Polewczak33}.
 In this paper,
 the calculation of $g_{2}(\vec{r}_{1},\vec{r}_{2}|n,\beta ;t)$ is applied to the method of collective Yukhnovskii variables, with the allocation of the frame of reference,
 which is a system of particles with a short-range nature of the interaction.
 In addition,
 it is important to note that it is promising to use the Ornstein--Zernike equation,
 which depends on the time~\cite{Ramsh1,Eu,Eu1,Bra} to calculate $g_{2}(\vec{r}_{1},\vec{r}_{2}|n,\beta ;t)$.

 In the following,
 we consider in detail the kinetic equation for a nonequilibrium one-particle distribution function with collision integral in the pair collision integral with a pair quasi-equilibrium coordinate distribution function in the Gaussian approximation (chaotic phases) in the Yukhnovskii method of collective variables.


%\bibliography{MyBasa2}

\end{document}